\documentclass[11pt,a4paper]{article} 
\usepackage{jinstpub}
\usepackage{lineno}
\notoc

\usepackage[caption=false]{subfig}
\usepackage{xcolor}

\makeatletter
\newlength{\sfp@hseplen}\newlength{\sfp@vseplen}
\define@cmdkey{subfigpos}[sfp@]{pos}[ul]{}
\define@cmdkey{subfigpos}[sfp@]{font}[\small]{}
\define@cmdkey{subfigpos}[sfp@]{vsep}[1.75\baselineskip]{\setlength{\sfp@vseplen}{\sfp@vsep}}
\define@cmdkey{subfigpos}[sfp@]{hsep}[10pt]{\setlength{\sfp@hseplen}{\sfp@hsep}}
\newcommand{\subfigimg}[3][,]{
	\setkeys{Gin,subfigpos}{pos,font,vsep,hsep,#1}
	\setbox1=\hbox{\includegraphics{#3}}
	\ifnum\pdfstrcmp{\sfp@pos}{ul}=0
	\leavevmode\rlap{\usebox1}
	\rlap{\hspace*{\sfp@hsep}\raisebox{\dimexpr\ht1-\sfp@vsep}{\sfp@font{#2}}}
	\phantom{\usebox1}
	\else\ifnum\pdfstrcmp{\sfp@pos}{ur}=0
	\leavevmode\usebox1
	\llap{\raisebox{\dimexpr\ht1-\sfp@vsep}{\sfp@font{#2}}\hspace*{\sfp@hsep}}
	\else\ifnum\pdfstrcmp{\sfp@pos}{lr}=0
	\leavevmode\usebox1
	\llap{\raisebox{\sfp@vsep}{\sfp@font{#2}}\hspace*{\sfp@hsep}}
	\else
	\leavevmode\rlap{\usebox1}
	\rlap{\hspace*{\sfp@hseplen}\raisebox{\sfp@vsep}{\sfp@font{#2}}}
	\phantom{\usebox1}
	\fi\fi\fi
}

\begin{document}

\title{High-resolution direct phase control in the spectral domain in ultrashort pulse lasers for pulse-shaping applications} 

\author[a,b,1]{Sean M Buczek\note{Corresponding author.}}
\author[b]{Gilbert W Collins}
\author[a]{Alexey Arefiev}
\author[b]{Mario J Manuel}
\affiliation[a]{University of California, San Diego\\9500 Gilman Dr, La Jolla, CA 92093, USA}
\affiliation[b]{General Atomics\\3550 General Atomics Court San Diego, CA 92121, USA}
\emailAdd{sbuczek@ucsd.edu}

\date{\today}

\abstract{
Ultrafast laser systems, those with a pulse duration on the order of picoseconds or less, have enabled advancements in a wide variety of fields. Of particular interest to this work, these laser systems are the key component to many High Energy Density (HED) physics experiments. Despite this, previous studies on the shape of the laser pulse within the HED community have focused primarily on pulse duration due to the relationship between pulse duration and peak intensity, while leaving the femtosecond scale structure of the pulse shape largely unstudied. To broaden the variety of potential pulses available for study, a method of reliably adjusting the pulse shape at the femtosecond scale using sub-nanometer resolution Direct Phase Control has been developed. This paper examines the capabilities of this new method compared to more commonplace dispersion-based pulse shaping methods. It also will detail the capabilities of the core algorithm driving this technique when used in conjunction with the WIZZLER and DAZZLER instruments that are common in high intensity laser labs. Performance of the method and instrumentation is examined using data taken with a single shot FROG system. Finally, some discussion is given to possible applications on how the Direct Phase Control pulse shaping technique will be implemented in the future.
}

\keywords{Plasma generation (laser-produced, RF, x ray-produced), Detector alignment and calibration methods (lasers, sources, particle-beams), Lasers}

\arxivnumber{2410.03135}

\maketitle 
\flushbottom

\section{\label{sec:introduction}Introduction}
Ultrafast lasers, meaning laser systems with full width at half maximum (FWHM) durations on the order of ps or less, were first shown to be physically possible in the 1960s \cite{Bass1968,Cubeddu1969} However, the advent of Chirped Pulse Amplification in 1985 \cite{Strickland1985} was the catalyst that truly pushed the field forward, allowing for high intensity lasers with FWHM durations measured in fs to be made. Today, these fs class lasers can be found in labs around the world, with peak powers ranging from only a few kW all the way into the PW range. \cite{Pennington1997,Jeong2014}

This wide range of power and increased availability has lead ultrafast lasers to be used in a wide variety of fields: from medical (LASIK\cite{Shen2019} and optogenetics\cite{Wanying2019}), to material processing \cite{Sugioka2014} and metrology\cite{Oh2005,Udem2002}, to more pure research focused endeavors such as High Energy Density (HED) physics\cite{Ditmire2004}. While each application brings with it a unique set of needs and challenges, this paper will focus mainly on applications for HED physics and the sensitivity these experiments have to the instantaneous field and intensity of the laser pulse, though the techniques and methods discussed may be useful to other fields and applications as well.

The difficulty in maintaining ultrashort pulses has increased as the pulse durations have shrank. This lead to the development of tools and systems to help control and stabilize the laser pulse. Critically to the work done in this paper, two instruments in particular have become fairly commonplace in ultrafast laser labs. The WIZZLER and the DAZZLER, both built by the company Fastlite, work hand in hand to maintain and adjust laser pulses\cite{Verluise2000}. Using a technique called Self-Referential Spectral Interferometry (SRSI), the WIZZLER is able to measure both the pulse spectral intensity and phase from a single shot\cite{Oksenhendler2010}, allowing it to produce a complete reconstruction of the shape of the pulse in time. Meanwhile, the DAZZLER is a type of Acousto-Optic Programmable Dispersive Filter (AOPDF)\cite{Tournois1997} capable of selectively adjusting the phase of specific wavelengths relative to each other. When used together, the two instruments allow for a feedback loop wherein the WIZZLER measure the spectrum and phase before telling the DAZZLER how to adjust the phase to minimize the pulse duration.

The DAZZLER makes use of an acousto-optic, birefringent crystal to control relative spectral phase. Birefringent crystals have the unique property that the index of refraction along one axis (referred to as the ordinary axis) is different than the index of refraction along another axis (the extraordinary axis). Combined with the property of acousto-optic crystals to diffract light based on the interaction between the sound and electromagnetic waves passing through them, it is possible to have two different wavelengths of light have two different optical path lengths despite going through the same physical length of crystal. This difference in optical path lengths creates a difference in phase between the two wavelengths relative to each other. Thus, with a carefully crafted acoustic pulse going through the crystal, it is possible to add or remove spectral phase differences across the full bandwidth of a laser pulse in a single pass.

The DAZZLER software calculates an acoustic pulse based on the requested phase input by the user. There are two ways the user can instruct the DAZZLER on how to adjust the phase. The first is by providing the DAZZLER with values for a 4th order dispersion polynomial, either through a settings file or the GUI interface for the DAZZLER software. This method has the benefit of being easily accessible and well understood, at the expense of being limited to phase profiles that can be described by a 4th order polynomial. The other option is Direct Phase Control (DPC), where an array of wavelengths, and the desired phase for each wavelength, are specified through a settings file. This is the same method that the WIZZLER uses during the feedback loop described earlier. DPC pulse shaping is more technically challenging as it changes the number of parameters the user is controlling from four to potentially several hundred, with each of these parameters having a different impact on the time varying amplitude of the pulse (commonly referred to as the pulse shape). However, DPC pulse shaping is also more powerful in terms of the variety of pulse shapes that it can create. 

As mentioned earlier, the feedback loop between the WIZZLER and DAZZLER will attempt to get the pulse FWHM duration to be as close as possible to the minimum duration the system is capable of producing, a state referred to as Fourier Transform Limited (FTL). This is the default because FTL is generally the desired operating state for many current applications using ultrafast lasers. Intuitively, having the shortest duration possible means compressing the same amount of energy per pulse into the shortest amount of time, giving the highest peak power. Within the field of HED physics high peak power has been sought after due to the increase in the coupling efficiency between the laser and particle beams. However, recent research\cite{Ziegler2021,Khanna2024,Permogorov2022,Tayyab2018,Souri2017,Kumar2019} has indicated that using specially shaped pulses may have benefits over having an FTL pulse. 

Notably, Zeigler\cite{Ziegler2021} used a DAZZLER to add varying amounts of 3rd order dispersion (TOD) to their pulse for a proton acceleration experiment. From this, they observed an increased total number of protons as well as increased maximum energy for the protons compared to their results with an FTL pulse. They looked to quantify the link between these values and the numerical amount of TOD but did not attempt to explain the role of the pulse shape itself on the outcomes they observed. Similarly, Permogorov\cite{Permogorov2022} and Tayyab\cite{Tayyab2018} both looked at the effect of chirp on protons accelerated from thin foils. Tayyab did discuss how the changed pulse shapes due to the chirp they introduced could have influenced their result but did not make the pulse shape the focus of their work, while Permogorov intentionally isolated and controlled for pulse shape, making sure their laser pulse was symmetric. While Souri\cite{Souri2017} and Kumar\cite{Kumar2019} both did look at the effect of the pulse shape using particle-in-cell simulations, no previous work could be found intentionally looking into the effect of laser pulse shape on laser driven plasma experimentally, creating a gap in the literature and leaving a potentially limited understanding of the underlying physics.

The work discussed herein begins to fill this gap in the present body of research by describing a method for implementing DPC over all spectral components, while also providing a reconstruction of the non-FTL pulse shape. This method relies on the initial spectrum of the pulse and its spectral phase. These two parameters were chosen because together they form a complete picture of the pulse and both can be measured simultaneously by instruments such as a WIZZLER,  FROG, or SPIDER. Given the completeness of the pulse reconstruction, such a reconstruction is able to easily look into nearly any numerical aspect of the laser pulse, including, but not limited to, the intensity as a function of time, the instantaneous electric or magnetic field of the pulse, the chirp, and any phase changes. By using a near-FTL measurement as an initial point, it is also possible for an algorithm to provide an on-shot reconstruction of the pulse shape as the DPC technique alters the spectral phase away from FTL, creating the opportunity for a robust diagnostic tool.

First, in section \ref{sec:theory} briefly goes over the mathematical basis for pulse shaping. Section \ref{sec:recon} then goes deeper into how the pulse reconstruction works while also explaining some of the capabilities and limitations of the hardware used for pulse shaping and testing. Section \ref{sec:experiment} presents experimental data showcasing the performance of DPC pulse shaping. Section \ref{sec:limits} discusses the biggest current limitations of the DPC pulse shaping technique and highlights some planned future developments. Finally, section \ref{sec:conclude} summarizes the work. 

\begin{figure}[]
	\hspace{-.12\textwidth}
	\subfloat[\vspace{-2\baselineskip}]{\label{fig:ftl}
		\subfigimg[width=.6\textwidth,pos=ul,hsep=39pt,vsep=17.5pt,font=\large\colorbox{gray!10}]{\subref{fig:ftl}}{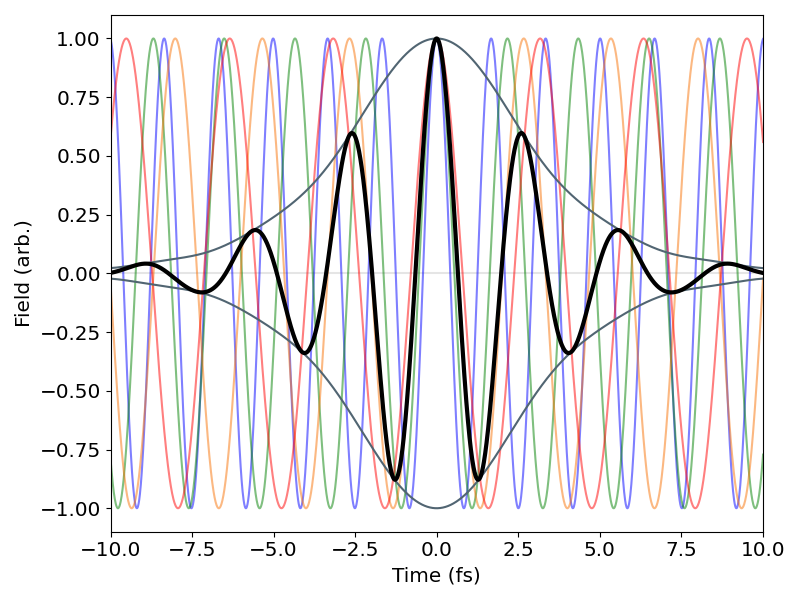}}
	\subfloat[\vspace{-2\baselineskip}]{\label{fig:split}
		\subfigimg[width=.6\textwidth,pos=ul,hsep=39pt,vsep=17.5pt,font=\large\colorbox{gray!10}]{\subref{fig:split}}{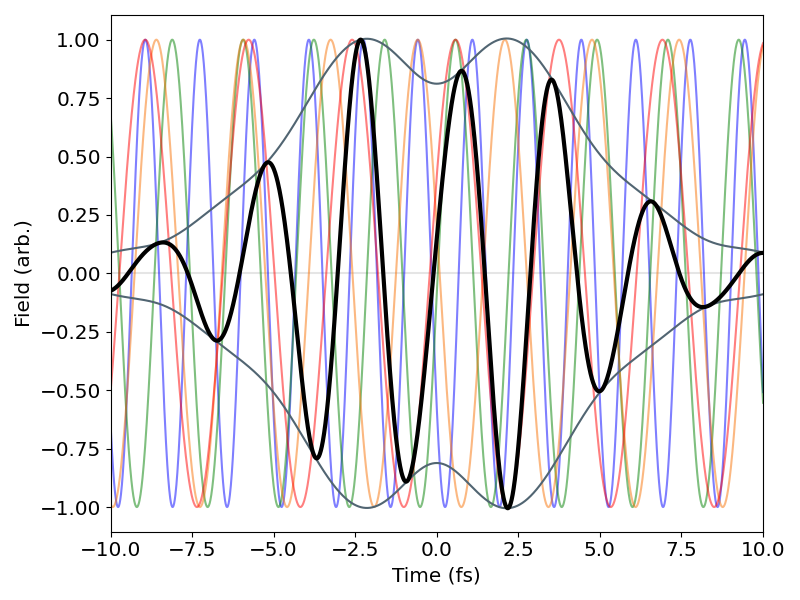}}
	\caption{\label{fig:superimpose} Examples of pulse envelope and field superimposed over a sample of frequencies that make up the pulses. The black line is the electric field of the pulse (with pulse envelope plotted in gray around it), while the multi-colored plots are four different frequencies out of the many that were used to create the overlaid pulse. The amplitude of all frequencies were normalized to themselves for the purposes of demonstration. Both figures use the same spectrum, with \protect\subref{fig:ftl} is at FTL while \protect\subref{fig:split} has the phase of alternating frequencies shifted such that their peaks occur either slightly earlier or later than t=0.}
\end{figure}

\section{\label{sec:theory}Mathematical Background} 
Mathematically, ultrashort laser pulses are nothing more than the superposition principle of light at work. If a large number waves, each with a unique frequency $\omega$, are superimposed with no restrictions imposed on the phase $\phi$ the resultant signal is purely noise since the waves will constructively and destructively interfere randomly. This resultant signal can be expressed as a sum of all the waveforms:

\begin{equation}
	\label{eq:fourier}
	E(t) = \sum_\omega A_\omega \exp{(-i\omega t +\phi)} + c.c.
\end{equation}

This is an application of the well-known Fourier series, written here in exponential form, with $A_\omega$ being the amplitude of each different frequency wave. If the phases of all the frequencies are all set to zero at any given time (for simplicity, time $t=0$) the resultant waveform will have one singular dominant peak with effectively zero signal outside this peak. This is the case for FTL laser pulses (figure \ref{fig:ftl}, pulse envelope showing one peak drawn in gray).

This dependence on phase is what allows the shape of the pulse to be controlled. By selectively shifting frequencies slightly away from zero phase, it is possible to change the overall shape of the pulse. Figure \ref{fig:split} shows exactly such a case. Each frequency has had the peak shifted in time by either plus or minus 0.58 fs from the center, alternating between plus and minus with each successive frequency. This is a non-physical system for the purposes of demonstrations. The spectrum being used is incredibly broad (600 nm, centered at 800 nm) and perfectly Gaussian across that bandwidth.

As can be seen in figure \ref{fig:split}, even this small shift of the peaks away from the center creates a symmetric double peaked pulse with minimal signal noise outside of the main pulse envelope. This type of pulse is very difficult to recreate using dispersion coefficients, as they struggle to create symmetric pulses with more than a single peak.

\begin{figure}[h]
	\centering
	\includegraphics[width=\linewidth]{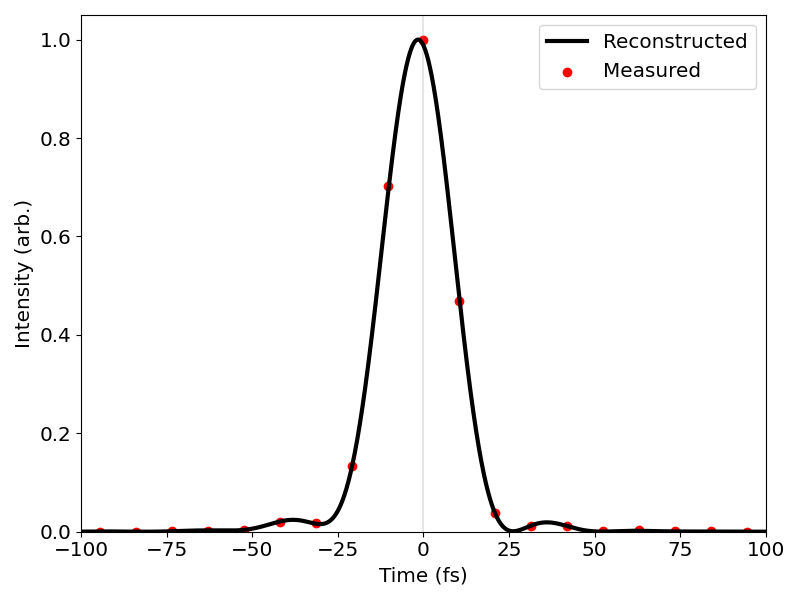}
	\caption{\label{fig:wizconfirm} A direct comparison of the pulse shape measured by the WIZZLER (red points) and reconstructed by the pulse reconstruction algorithm.}
\end{figure}

\section{\label{sec:recon}Pulse Reconstruction}
In order to numerically reconstruct the electric field of the laser pulse, the straightforward nature of the Fourier series is utilized. Using the spectral intensity and the spectral phase of the laser pulse it is possible to perform a Discrete Fourier Transform (DFT) on the electric fields of the spectrum. This allows the spectral intensity and phase to be retrieved at any time while still providing a full description of the pulse. This method also allows for the flexible implementation of diagnostic features, such as the ability to input changes to the phase and have pulse parameters be quickly recalculated.

This process for pulse reconstruction is highly accurate and trustworthy, being identical to well-established theory. On top of that,  results of the reconstruction algorithm developed for this method were compared to intensity measurements that the WIZZLER took for a near-FTL pulse and found the results matched to within rounding error (0.1 fs or less difference in FWHM duration). Qualitatively, the measured and reconstructed pulse shapes are clearly equivalent (figure \ref{fig:wizconfirm}). The data for this figure was taken using the GALADRIEL laser system\cite{Collins2024} located at General Atomics in San Diego. This is a Ti:Sapphire laser with a center wavelength of 800 nm and a FWHM bandwidth of up to 80 nm.

While this method of pulse shaping and reconstruction is independent of any particular system, it was designed with the intent of being used in partnership with a DAZZLER. To that end, it is worth briefly discussing the performance, and limitations, of the DAZZLER. The specific parameters of concern are the smallest spectral division the DAZZLER can effect, how much the DAZZLER can change the phase of each spectral step relative to the previous one, and how often can changes to the phase be sent to the DAZZLER such that those changes reliably affect the next laser pulse.

The first of these, what is essentially the spectral resolution of the DAZZLER, can be calculated based on system parameters. This relationship is given in the DAZZLER documentation\cite{dazzlermanual} as:
\begin{equation}
	\label{eq:specres}
	\delta\lambda_{1/2} = \frac{0.8\lambda^2}{\delta nL} \approx 8.9\frac{\lambda^2}{L}
\end{equation}
Where $L$ is the length of the crystal in the DAZZLER system, $\delta n$ is the difference in index of refraction between the ordinary and extraordinary axes, and $\lambda$ is the center wavelength of the spectrum. The spectral resolution can also be found by looking at the settings files for the DAZZLER. For GALADRIEL, this value is 0.153 nm.

\begin{figure}
	\hspace{-.12\textwidth}
	\subfloat[\vspace{-2\baselineskip}]{\label{fig:spolyphase}
		\subfigimg[width=.6\textwidth,pos=ul,hsep=28pt,vsep=17.5pt,font=\large\colorbox{gray!10}]{\subref{fig:spolyphase}}{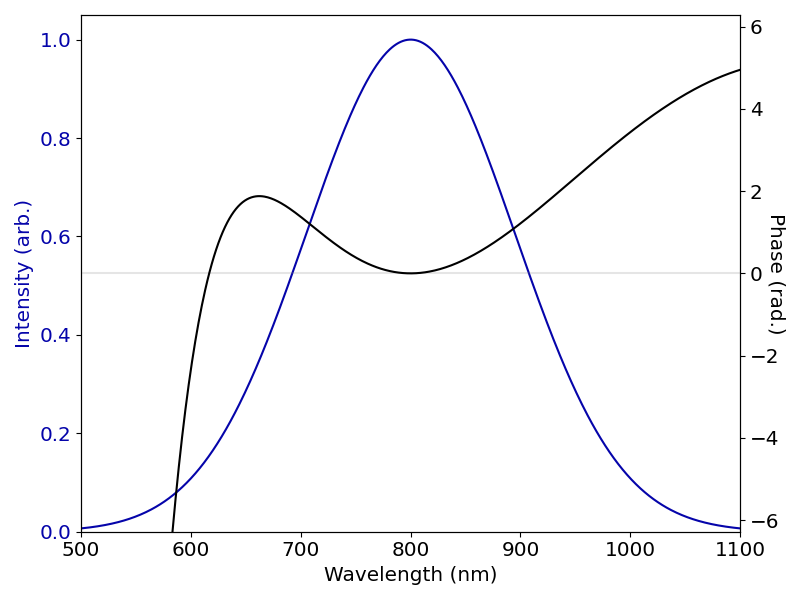}}
	\subfloat[\vspace{-2\baselineskip}]{\label{fig:spolyshape}
		\subfigimg[width=.6\textwidth,pos=ul,hsep=28pt,vsep=17.5pt,font=\large\colorbox{gray!10}]{\subref{fig:spolyshape}}{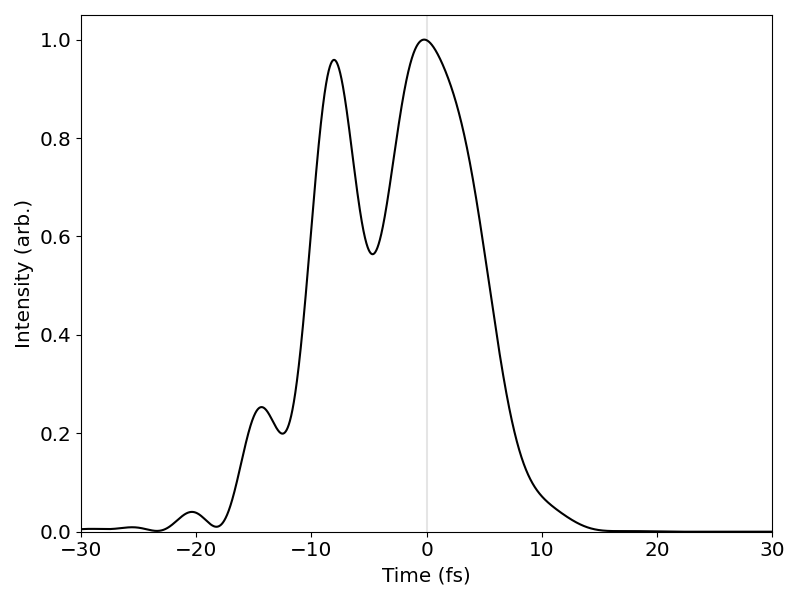}}
	\vspace{-\baselineskip}
	
	\hspace{-.12\textwidth}
	\subfloat[\vspace{-2\baselineskip}]{\label{fig:sdirephase}
		\subfigimg[width=.6\textwidth,pos=ul,hsep=28pt,vsep=17.5pt,font=\large\colorbox{gray!10}]{\subref{fig:sdirephase}}{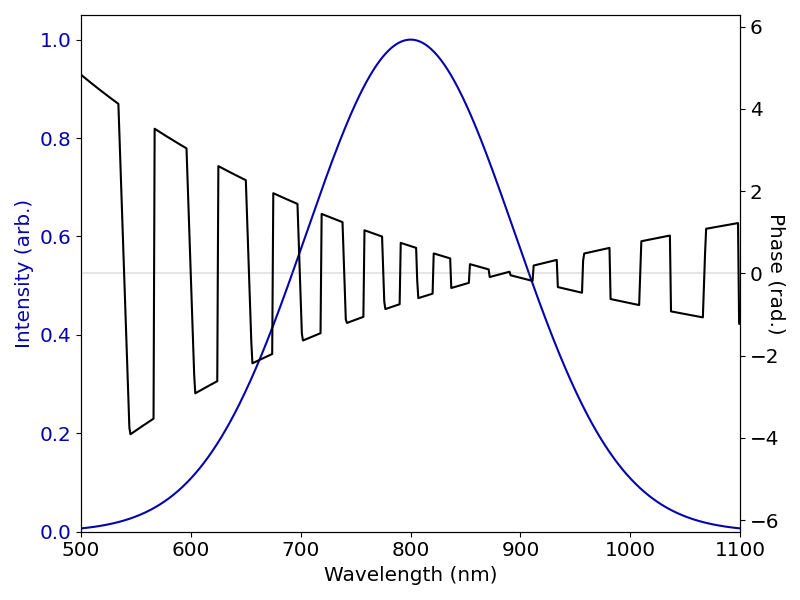}}
	\subfloat[\vspace{-2\baselineskip}]{\label{fig:sdireshape}
		\subfigimg[width=.6\textwidth,pos=ul,hsep=28pt,vsep=17.5pt,font=\large\colorbox{gray!10}]{\subref{fig:sdireshape}}{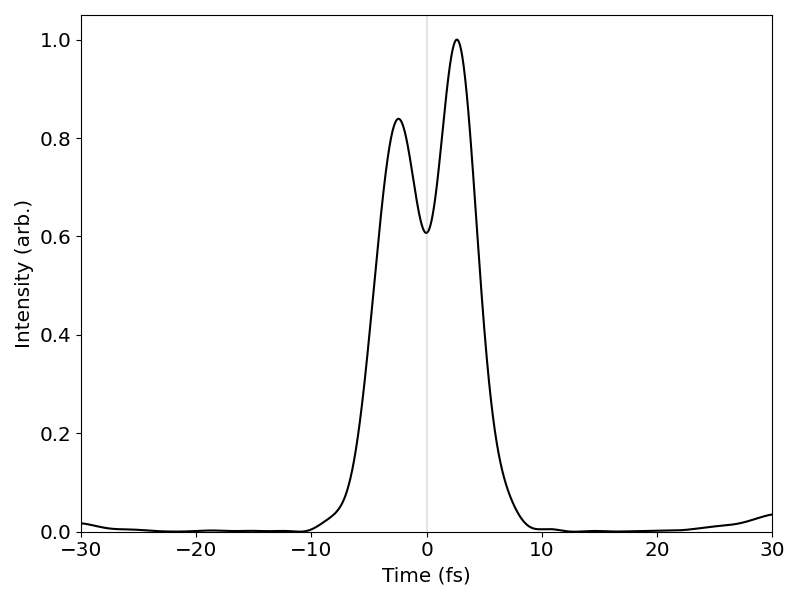}}
	
	\caption{\label{fig:control} Examples of the capabilities of pulse shaping using dispersion polynomials compared to DPC. \protect\subref{fig:spolyphase} shows the spectrum and phase created using a dispersion polynomial while \protect\subref{fig:spolyshape} is the resultant pulse. Similarly, \protect\subref{fig:sdirephase} shows the spectrum and phase created using the DPC technique with \protect\subref{fig:sdireshape} being the resultant pulse. Note that the spectrum used for both \protect\subref{fig:spolyphase} and \protect\subref{fig:sdirephase} is the same, only the phase is different.}	
\end{figure}

The second parameter, the maximum phase change per spectral step, had to be found experimentally. The DAZZLER has limits built into its software that will prevent it from trying to output acoustic waveforms that it is unable to create. Based on those, the minimum number of steps needed to change the phase by $\pi$ radian was found to be 0.61 nm. In other words, when 800 nm was set to be zero phase, the next closest wavelength that could have a phase of $\pi$ radian was 800.61 nm, with the steps in between them set as linear steps between zero and $\pi$. This indicated that the maximum shift per division is roughly $\pi/4$ radian. It is possible this value is also system specific. 

Similarly, testing on the GALADRIEL system has shown that the fastest frequency that the DAZZLER reliably change setpoints to have them accurately reflected in the next pulse is at 2 Hz. Trying to change setpoints more often than that would result in pulses occasionally not updating immediately, which gives incorrect data regarding pulse shape. The occurrence of this updating delay was not consistent, but would happen more often with more rapid attempts to send changes to the DAZZLER. In other words, it occurred on the order of once every hundred shots at 3 Hz, but on the order of once every 20 shots at 4 Hz, with no clear pattern that could be used to predict the next update delay.

Working within these parameters, demonstrating the difference in pulse shape control when using DPC as compared to using dispersion polynomials is straightforward. Using the same synthetic dataset used in figure \ref{fig:superimpose} as a baseline, the phase was manipulated either with dispersion polynomials (figure \ref{fig:spolyphase}) or DPC (figure \ref{fig:sdirephase}). The goal was to transform the initial FTL pulse into pulse with two peaks that is as symmetric as possible. For the purposes of this work, these adjustments were done manually, while a future implementation of this technique will leverage machine learning to assist in building pulse shapes tailored at the fs level. For the dispersion polynomial control, 2nd through 4th order dispersion coefficients were used as those are what affect the pulse shape and are available in the DAZZLER GUI. To maintain parity with what can actually be done experimentally, the phase for the DPC tests was not shifted by more than $\pi/4$ rad/step. Figures \ref{fig:spolyshape} and \ref{fig:sdireshape} show the pulse shapes that resulted from the respective phase changes.

\begin{figure}[]
	\hspace{-.12\textwidth}
	\subfloat[\vspace{-2\baselineskip}]{\label{fig:2ndshape}
		\subfigimg[width=.6\textwidth,pos=ul,hsep=28pt,vsep=17.5pt,font=\large\colorbox{gray!10}]{\subref{fig:2ndshape}}{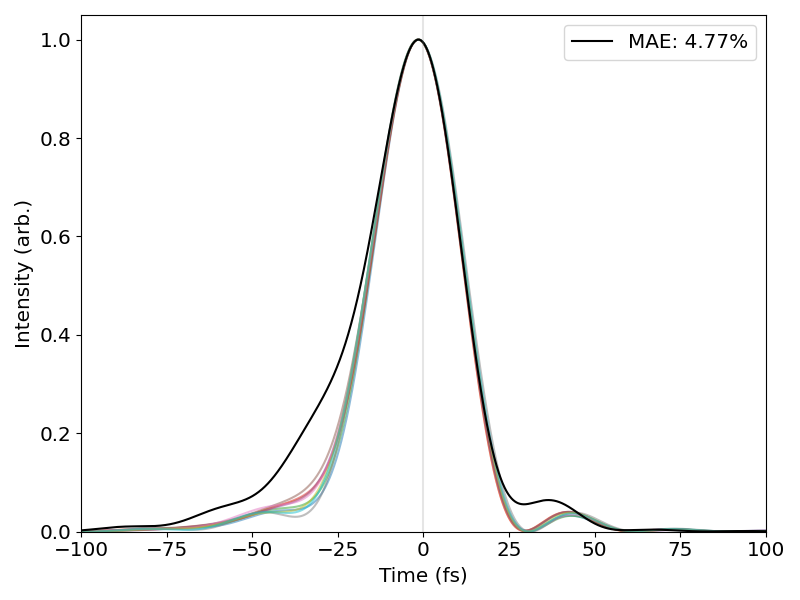}}
	\subfloat[\vspace{-2\baselineskip}]{\label{fig:3rdshape}
		\subfigimg[width=.6\textwidth,pos=ul,hsep=28pt,vsep=17.5pt,font=\large\colorbox{gray!10}]{\subref{fig:3rdshape}}{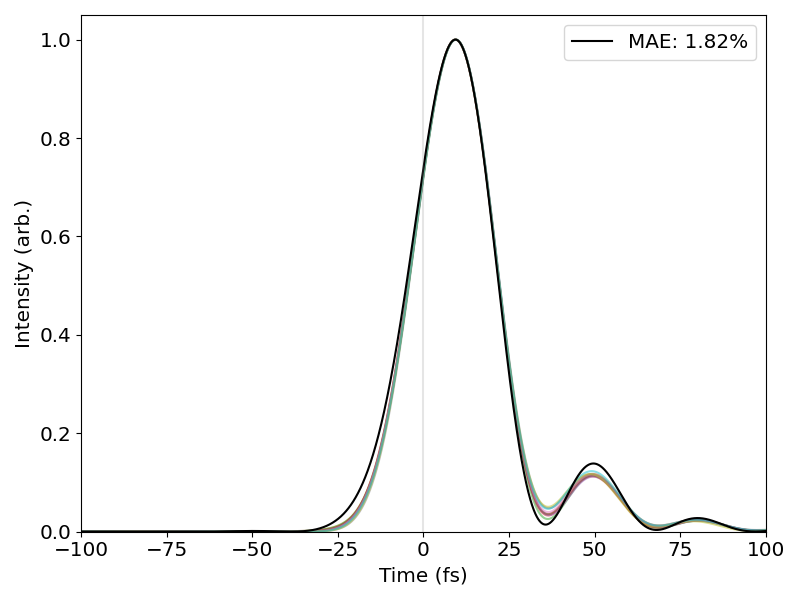}}
	\vspace{-\baselineskip}
	
	\hspace{-.12\textwidth}
	\subfloat[\vspace{-2\baselineskip}]{\label{fig:2ndspec}
		\subfigimg[width=.6\textwidth,pos=ul,hsep=28pt,vsep=17.5pt,font=\large\colorbox{gray!10}]{\subref{fig:2ndspec}}{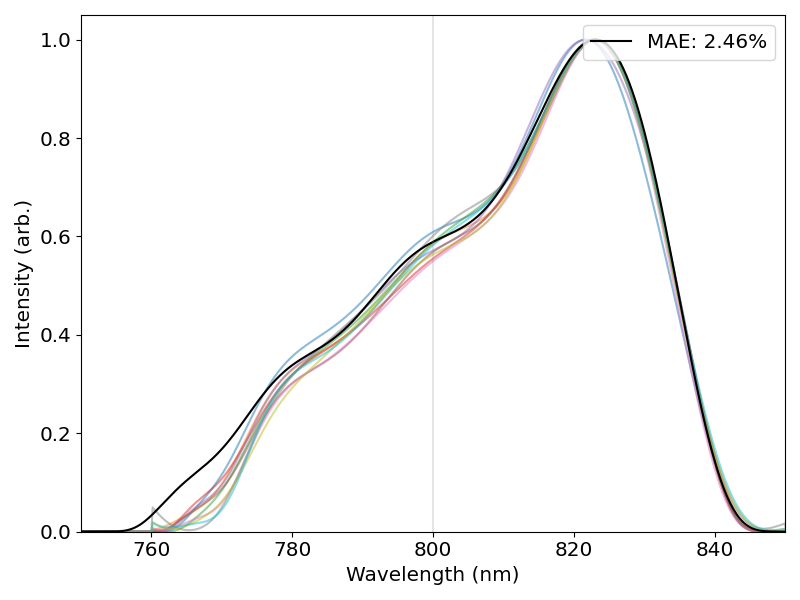}}
	\subfloat[\vspace{-2\baselineskip}]{\label{fig:3rdspec}
		\subfigimg[width=.6\textwidth,pos=ul,hsep=28pt,vsep=17.5pt,font=\large\colorbox{gray!10}]{\subref{fig:3rdspec}}{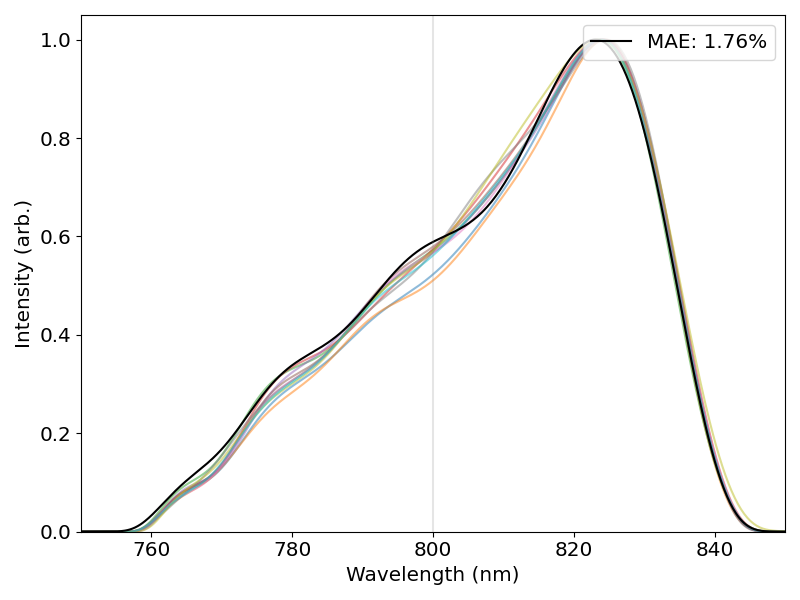}}
	\vspace{-\baselineskip}
	
	\hspace{-.12\textwidth}
	\subfloat[\vspace{-2\baselineskip}]{\label{fig:2ndphase}
		\subfigimg[width=.6\textwidth,pos=ul,hsep=28pt,vsep=17.5pt,font=\large\colorbox{gray!10}]{\subref{fig:2ndphase}}{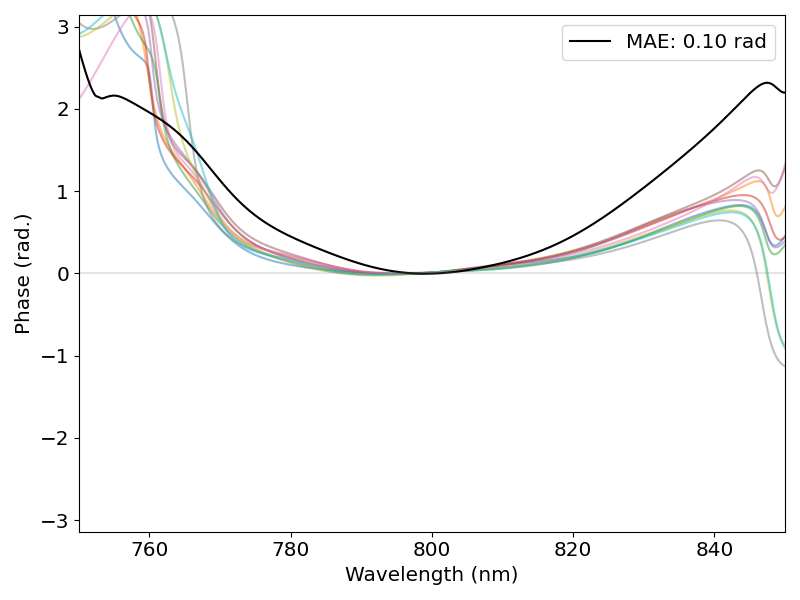}}
	\subfloat[\vspace{-2\baselineskip}]{\label{fig:3rdphase}
		\subfigimg[width=.6\textwidth,pos=ul,hsep=28pt,vsep=17.5pt,font=\large\colorbox{gray!10}]{\subref{fig:3rdphase}}{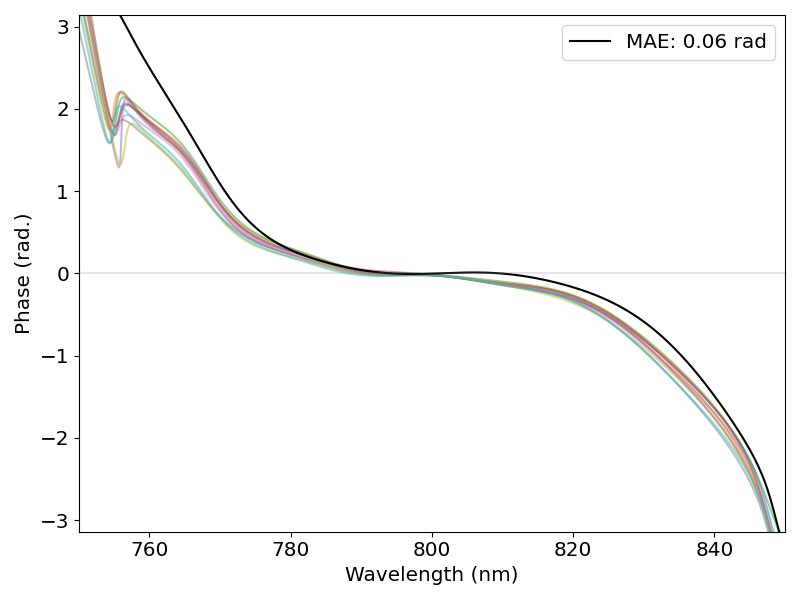}}
	
	\caption{\label{fig:disper} The pulse shape (\protect\subref{fig:2ndshape} and \protect\subref{fig:3rdshape}), spectrum (\protect\subref{fig:2ndspec} and \protect\subref{fig:3rdspec}), and spectral phase (\protect\subref{fig:2ndphase} and \protect\subref{fig:3rdphase}) for two different amounts of added dispersion: \protect\subref{fig:2ndshape}, \protect\subref{fig:2ndspec}, \protect\subref{fig:2ndphase} 240 fs$^2$ of 2nd order dispersion and \protect\subref{fig:3rdshape}, \protect\subref{fig:3rdspec}, \protect\subref{fig:3rdphase} 7500 fs$^3$ of 3rd order dispersion. In all cases, the black lines are the theoretical pulse, while the colored lines represent a sampling of measured pulses. The theoretical spectra are the same, as they are the measured spectrum of the pulse at best compression from that day.}	
\end{figure}

\section{\label{sec:experiment}Experimental Verification}
In order for the DPC pulse reconstruction to act as an on-shot diagnostic, it is necessary for the pulse shape to be accurately reconstructed when diagnostics such as the WIZZLER are unable to be used. While baseline data can be easily recorded, certain experimental setups do not allow for extra diagnostics to be added close to the target. To enable these reconstructions, the core algorithm was designed to be flexible enough to recalculate the pulse reconstruction when provided new inputs such as a new spectral intensity measurements, changes to the provided spectral phase array, or specified dispersion coefficients. This flexibility enables the user to adjust the resultant pulse based on optical medium they know the laser will have to travel through, such as windows or dielectric coatings and still have an accurate reconstruction. 

To show that the DPC algorithm dispersion calculations accurately reflect changes to the laser pulse, a direct comparison with measured data was needed. This can be done by starting with an initial laser pulse optimized to be as close to FTL as possible using the standard WIZZLER-DAZZLER feedback loop. The DAZZLER GUI can then be used to introduce specific, known amounts of 2nd, 3rd, and 4th order dispersion to the pulse, while the initial and dispersion-added pulses are recorded using both a WIZZLER and a commercially available single-shot FROG system from Swamp Optics. Feeding the same initial pulse and known dispersion values into the reconstruction algorithm will produce the expected pulse. For this testing, figure \ref{fig:disper} shows the plots of pulse shape, spectrum, and spectral phase for 2nd and 3rd order dispersion. The black curve being the reconstructed pulse and the colored lines being a sampling of the data taken by the WIZZLER. Shot-to-shot measurement variation is illustrated by the different colored lines in figure \ref{fig:disper}. Comparing the predicted pulse to each of these sampled pulses, the mean absolute error for the 2nd, 3rd, and 4th order dispersion was found to be $4.77\%$, $1.82\%$, and $4.87\%$ respectively. Similarly, the mean absolute error between the predicted pulse and those measured by the FROG was $3.49\%$, $5.36\%$, and $5.92\%$ for the 2nd, 3rd, and 4th order dispersion respectively. Notably, while the spectrum does not appear to be affected by this dispersion based pulse shaping in any meaningful way, the measured phases (figures \ref{fig:2ndphase} and \ref{fig:3rdphase}) do not match precisely what is expected. This deviation may explain the error in the reconstructed pulse shapes (figures \ref{fig:2ndshape} and \ref{fig:3rdshape}).

\begin{figure}[]
	\hspace{-.12\textwidth}
	\subfloat[\vspace{-2\baselineskip}]{\label{fig:stepshape}
		\subfigimg[width=.6\textwidth,pos=ul,hsep=28pt,vsep=17.5pt,font=\large\colorbox{gray!10}]{\subref{fig:stepshape}}{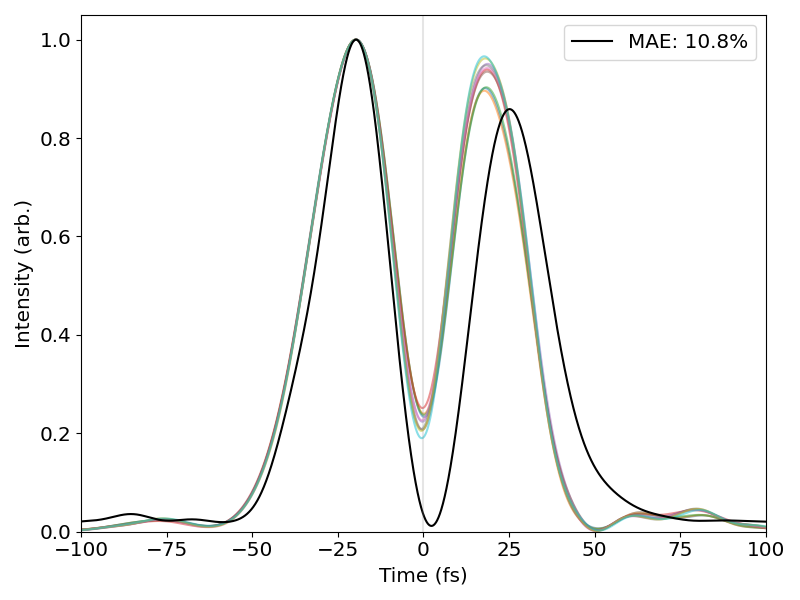}}
	\subfloat[\vspace{-2\baselineskip}]{\label{fig:sawshape}
		\subfigimg[width=.6\textwidth,pos=ul,hsep=28pt,vsep=17.5pt,font=\large\colorbox{gray!10}]{\subref{fig:sawshape}}{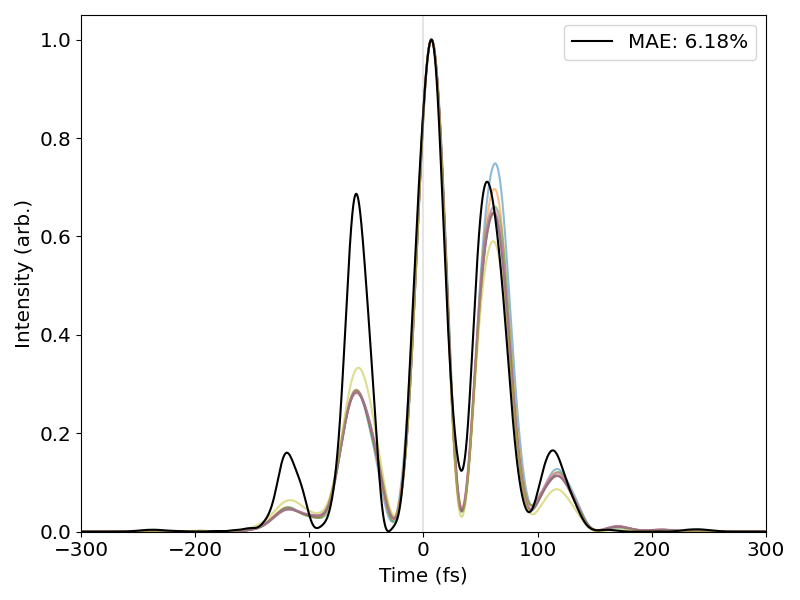}}
	\vspace{-\baselineskip}
	
	\hspace{-.12\textwidth}
	\subfloat[\vspace{-2\baselineskip}]{\label{fig:stepspec}
		\subfigimg[width=.6\textwidth,pos=ul,hsep=28pt,vsep=17.5pt,font=\large\colorbox{gray!10}]{\subref{fig:stepspec}}{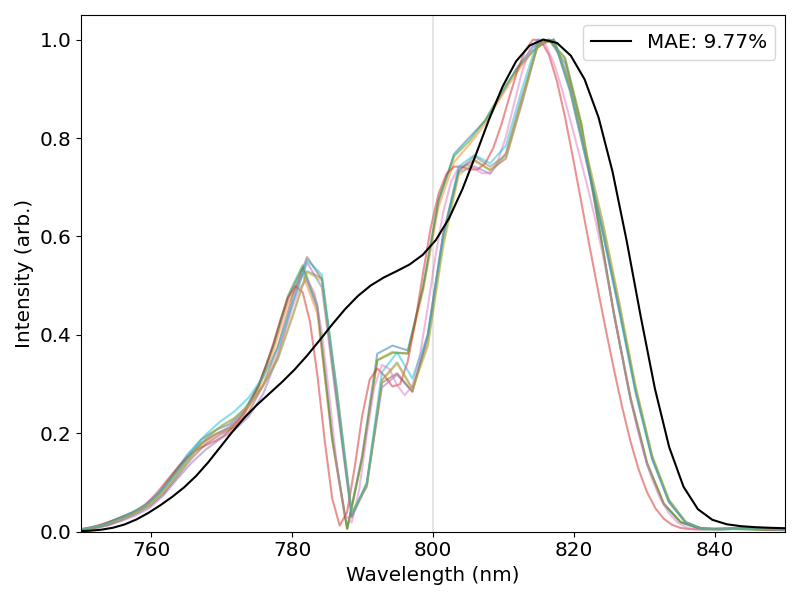}}
	\subfloat[\vspace{-2\baselineskip}]{\label{fig:sawspec}
		\subfigimg[width=.6\textwidth,pos=ul,hsep=28pt,vsep=17.5pt,font=\large\colorbox{gray!10}]{\subref{fig:sawspec}}{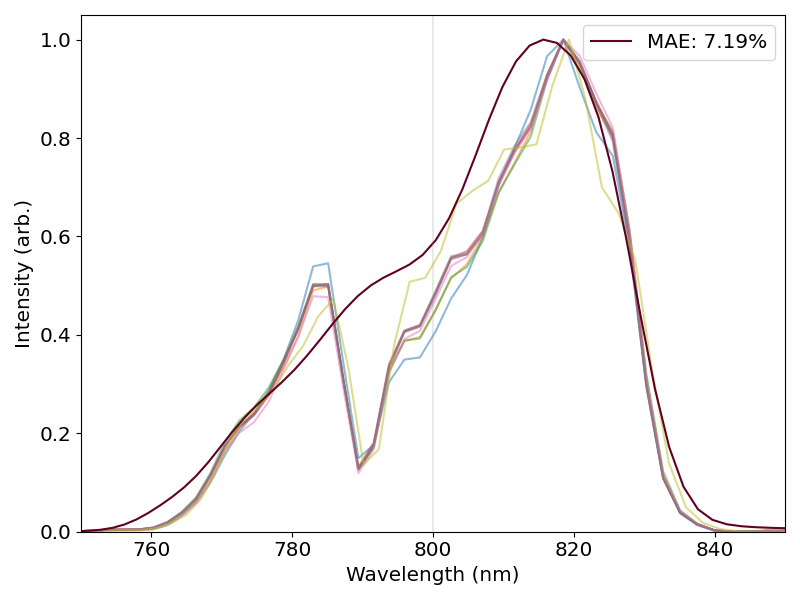}}
	\vspace{-\baselineskip}
	
	\hspace{-.12\textwidth}
	\subfloat[\vspace{-2\baselineskip}]{\label{fig:stepphase}
		\subfigimg[width=.6\textwidth,pos=ul,hsep=28pt,vsep=17.5pt,font=\large\colorbox{gray!10}]{\subref{fig:stepphase}}{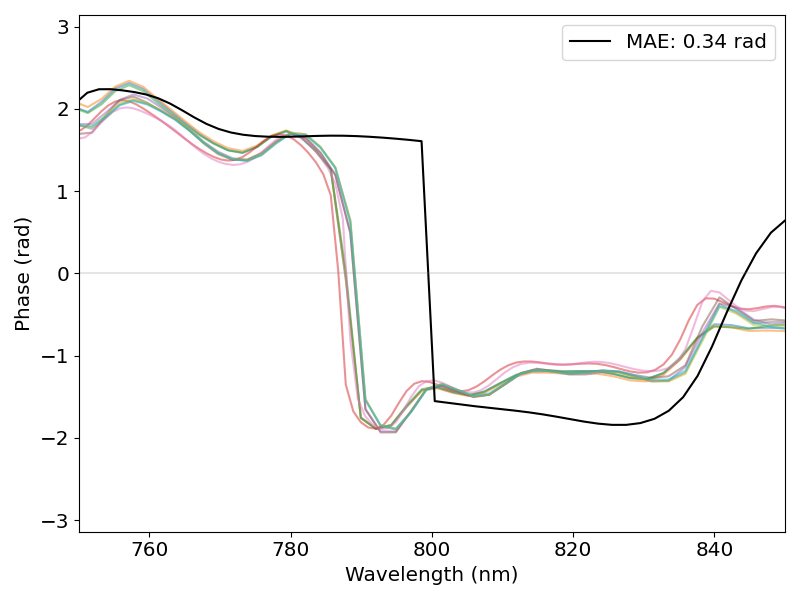}}
	\subfloat[\vspace{-2\baselineskip}]{\label{fig:sawphase}
		\subfigimg[width=.6\textwidth,pos=ul,hsep=28pt,vsep=17.5pt,font=\large\colorbox{gray!10}]{\subref{fig:sawphase}}{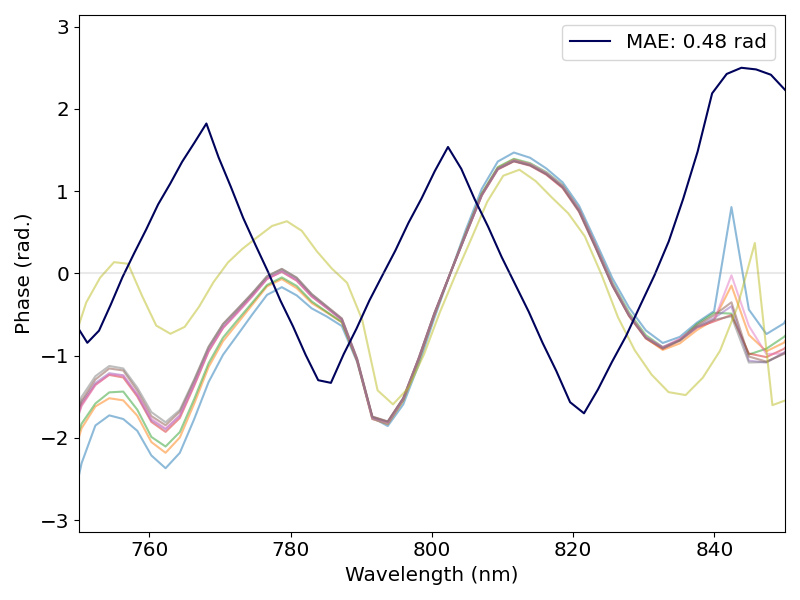}}
	
	\caption{\label{fig:arbphase} The pulse shape (\protect\subref{fig:stepshape} and \protect\subref{fig:sawshape}), spectrum (\protect\subref{fig:stepspec} and \protect\subref{fig:sawspec}), and spectral phase (\protect\subref{fig:stepphase} and \protect\subref{fig:sawphase}) of two different shaped pulses. In all cases, the black lines are the theoretical pulse, while the colored lines represent a sampling of measured pulses. The theoretical spectra are the same, as they are the measured spectrum of the pulse at best compression from that day.}	
\end{figure}

Aiming to use the same method of comparison, the DAZZLER was provided with several arrays of specifically selected phases such that a dispersion polynomial wouldn’t be able to create, notably one that included a step function style shift in the phase (figure \ref{fig:stepphase}) and one that used a triangular wave function (figure \ref{fig:sawphase}). This was to show that the DAZZLER is capable of creating these arbitrary phase profiles and that it is possible to accurately predict the changes to the pulse shape that result from the DPC technique. The FROG was used to record the initial and resultant pulses. The WIZZLER was not included in this round of testing as it is limited to measuring pulses that are near-FTL while the shapes tested were significantly altered enough that the WIZZLER could not consistently measure them. The initial pulse spectrum and phase, as well as the specially adjusted phase array, were then used as inputs to the reconstruction algorithm. The resulting pulse shapes are plotted in figures \ref{fig:stepshape} and \ref{fig:sawshape}. Comparing the predicted pulse shape to each of the sampled pulses, the mean absolute error is $10.8\%$ and $6.18\%$ for the step function and triangular wave, respectively.

The measured and theoretical spectral phase were also compared to see how accurately the DAZZLER is able to reconstruct a requested phase array. The mean absolute error was below 0.5 radian for both the step function and triangular wave shapes. However, it is qualitatively clear that there is a wavelength shift between the major features in the theoretical (requested) phase and the measured phase. This is most obvious in the step function, where the requested step was centered at 800 nm, while the measured step occurred around roughly 790 nm. A similar shift was seen for all of the phase shapes that were tested. The source of this shift has not yet been identified, but it is believed to be related to instrument settings and not inherently an effect of the DPC technique.

Similarly, the spectrum of the laser was affected by the pulse shaping. The spectrum used for all the theoretical pulses was the one measured for the laser at best compression during laser start-up the same day as the measured data was taken. The large departures from this initial spectrum were verified with an external spectrometer and thus are actual physical effects, not measurement or algorithmic reconstruction errors. These departures occur at the same wavelengths that major changes to the phase happen. More significant phase changes, such as were created using the DPC technique and seen in figures \ref{fig:stepspec} and \ref{fig:sawspec}, resulted in more significant spectrum changes compared to the more gentle phase changes seen with dispersion based pulse shaping (see figure \ref{fig:disper}). The differences between the expected and measured spectra in combination with the wavelength shift in the phase explain the error in the DPC pulse shape.

\section{\label{sec:limits}Limitations and Future Work}
In contrast to the more established dispersion polynomial based pulse shaping, the DPC technique still has several unknowns that require further study. This is one of the two major limitations with the technique as it currently is. Future studies are planned to include how DPC pulse shaping interacts with other instruments involved in laser pulse generation, such as amplifiers, compressors, and non-linear media. These studies should reveal why the DPC technique effects the laser spectrum in the way that is seen in this work as well as why the wavelength shift in the phase array happens and how to correct it. Understanding these effects will greatly improve the accuracy of the DPC technique in the future.

The other major limitation is that the current implementation of the DPC technique is a very manually intensive process. All phase arrays and pulse shapes shown in this work were crafted through a guess-and-check process until viable setpoints were found. During this manual process it was found that the pulse shape is even more sensitive to changes in spectral phase than anticipated and could very rapidly change. Despite these limitations, DPC pulse shaping has demonstrated the capability of finely adjusting the pulse shape away from FTL and will unlock the bespoke laser pulses needed to probe yet unexplored phenomena.

There are plans in place to mitigate both of these limitations. The first are the previously mentioned studies to better understand how the DPC pulse shaping technique impacts other parts of the system and if it is possible to compensate the effect it has on the spectrum with another device. The second limitation is being addressed with the development of a software tool that will enable users to specify the desired pulse shape. This tool is being developed as part of a long-term effort to integrate the DPC technique into the suite of diagnostics currently available on GALADRIEL for use as an on-shot pulse shape diagnostic.

\section{\label{sec:conclude}Conclusion}
The ability of this technique to accurately reconstruct precise pulse shape enables the creation of a reliable on-shot diagnostic tool. It is able to predict pulse shapes to within $<6\%$ error, with future improvements likely to reduce this error further, and will enable better on-shot information about the laser. This will better inform the physics of what is happening in the laser plasma interaction.  

 The implementation of DPC pulse shaping on GALADRIEL is still being developed. It is intended to be used for both on-shot diagnostics of the laser pulse as well as part of a control loop for the laser, helping improve stability for tests that involve pulse shaping. As a part of this control loop, it will eventually enable system users to specify arbitrary pulse shapes without the user needing to know the exact spectrum and phase needed to create them.

These capabilities rely on the assumption that the laser system is stable over some known amount of time. All lasers systems are known to have spectral drift over the course of a day of operation, but initial testing has shown that the GALADRIEL system remains close to an initial point over timescales on the order of single hours. Given that retaking baseline data and recalibrating takes only a couple of minutes, with possible upgrades to the lab planned to reduce that time even further, the amount of uptime is sufficient to justify the use of this method as a diagnostic given the significant advantages it provides.

\section{\label{sec:ack}Acknowledgements}
S.B., M.M., and A.A. were supported in part by the National Science Foundation–Czech Science Foundation partnership (NSF Grant No. PHY-2206777). GALADRIEL access provided by General Atomics internal research and development.

\bibliographystyle{JHEP}
\bibliography{InstPaperRef.bib}

\end{document}